\documentclass[journal]{IEEEtran}

\usepackage{amsbsy}
\usepackage{amsmath}
\usepackage{mathtools}
\usepackage{amsfonts} % Needed for blackboard bold, etc
\usepackage{amssymb}  % a superset of amsfonts
\usepackage{amsthm}   % for fancy theorem styles
\usepackage{amsxtra}
\usepackage{bm}
\usepackage{multirow}
\usepackage{nomencl}
\usepackage{color,colortbl}
\usepackage{dsfont}
\usepackage{enumerate}
\usepackage{fancybox}
\usepackage{fancyhdr}
\usepackage{graphics}
\usepackage{graphicx}
\usepackage{latexsym}
\usepackage{subfig}
\graphicspath{{./images/}}
\usepackage{algorithmic}

\definecolor{Dblue}{rgb}{0,0,1}
\definecolor{Dbrown}{rgb}{0.59,0.4,0}
\definecolor{Dred}{rgb}{0.64,0,0}
\definecolor{Dgreen}{rgb}{0,0.4,0}
\def\real{\mathbb R}
\def \bs {\boldsymbol}
\def \mr {\mathrm}
\def \tb {\textbf}

\def \gr {\nabla}

\newtheorem{theorem}{Theorem}[section]
\newtheorem{lemma}{Lemma}[section]

\makenomenclature
\begin{document}

\title{Direction of Arrival Estimation Using Co-prime Arrays: A Super Resolution Viewpoint}
\author{Zhao~Tan~\IEEEmembership{Student Member,~IEEE}, ~Yonina~C.~Eldar~\IEEEmembership{Fellow,~IEEE},
        and~Arye~Nehorai~\IEEEmembership{Fellow,~IEEE}
\thanks{Z. Tan and A. Nehorai are with the Preston M. Green Department of Electrical and Systems Engineering Department, Washington University in St. Louis, St. Louis,
MO, 63130 USA. E-mail: \{tanz, nehorai\}@ese.wustl.edu.}% <-this % stops a space
\thanks{Y. C. Eldar is with the Department of Electrical Engineering, Technion---Israel Institute of Technology, Haifa 32000, Isreal. E-mail: yonina@ee.technion.ac.il.}
\thanks{The work of Z. Tan and A. Nehorai was supported by the AFOSR Grant FA9550-11-1-0210, and ONR Grant N000141310050. The work of Y. C. Eldar was supported in part by the Israel Science Foundation under Grant no. 170/10, in part by the Ollendorf Foundation, and in part by a Magneton from the Israel Ministry of Industry and Trade.}}% <-this % stops a space}

\maketitle

\begin{abstract}
We consider the problem of direction of arrival (DOA) estimation using a newly proposed structure of non-uniform linear arrays, referred to as co-prime arrays, in this paper. By exploiting the second order statistical information of the received signals, co-prime arrays exhibit $O(MN)$ degrees of freedom with only $M+N$ sensors. A sparsity based recovery method is proposed to fully utilize these degrees of freedom. Unlike traditional sparse recovery methods, the proposed method is based on the developing theory of super resolution, which considers a continuous range of possible sources instead of discretizing this range into a discrete grid. With this approach, off-grid effects inherited in traditional sparse recovery can be neglected, thus improving the accuracy of DOA estimation. In this paper we show that in the noiseless case one can  theoretically detect up to $\frac{MN}{2}$ sources with only $2M+N$ sensors. The noise statistics of co-prime arrays are also analyzed to demonstrate the robustness of the proposed optimization scheme. A source number detection method is presented based on the spectrum reconstructed from the sparse method. By extensive numerical examples, we show the superiority of the proposed method in terms of DOA estimation accuracy, degrees of freedom, and resolution ability compared with previous methods, such as MUSIC with spatial smoothing and the discrete sparse recovery method.
\end{abstract}
  

\section{Introduction}
In the last few decades, research on direction of arrival (DOA) estimation using array processing has focused primarily on uniform linear arrays (ULA) \cite{ULA}. It is well known that by implementing a ULA with $N$ sensors, the number of sources that can be resolved by MUSIC-like algorithms is $N-1$ \cite{music}. New geometries \cite{nested, coprime} of non-uniform linear arrays have been recently proposed to increase the degrees of freedom of the array by studying the covariance matrix of the received signals among different sensors. By vectorizing the covariance matrix, the system model can be viewed as a virtual array with a wider aperture. In \cite{nested}, a nested array structure was proposed to increase the degrees of freedom from $O(N)$ to $O(N^2)$, with only $O(N)$ sensors. However, some of the sensors in the nested array structure are closely located, which leads to mutual coupling among these sensors. To overcome this shortcoming, co-prime arrays were proposed in \cite{coprime}, and it was shown that by using $O(M+N)$ sensors, this structure can achieve $O(MN)$ degrees of freedom. In this paper we will focus on co-prime arrays.

The increased degrees of freedom provided by the co-prime structure can be utilized to improve DOA estimation. Two main methodologies have been proposed. One is subspace methods, such as the MUSIC algorithm. In \cite{comusic}, a spatial smoothing technique was implemented prior to the application of MUSIC. The authors showed that an increased number of sources can be detected by the co-prime arrays. However, the application of spatial smoothing reduces of the obtained virtual array aperture \cite{cscoprime}. The second methodology uses sparsity based recovery to overcome this disadvantages of subspace methods \cite{cscoprime}\nocite{corr}\nocite{corrIm}-\cite{corrGS}. Traditional sparsity based recovery discretizes the range of interest into a grid. The assumption made by sparsity methods is that all sources are located exactly at the grid points. However, off-grid targets can lead to mismatches in the model and deteriorate the performance of sparse recovery significantly \cite{sen_mismatch}. In \cite{Zhao13,Xie12} the grid mismatches were estimated simultaneously with the original signal, and they showed that by considering grid mismatches one can achieve a better sparse recovery performance than the traditional sparse recovery methods. In \cite{ZT13_SPL}, the joint sparsity between the original signal and the mismatch was exploited during the DOA estimation for co-prime arrays. Due to the first order approximation used in \cite{ZT13_SPL}, the estimation performance is still limited by the higher order modeling mismatch. 

To overcome this difficulty of traditional sparsity based methods, a recent developed mathematical theory of super resolution \cite{SuperRes,SuperNoise} is utilized in this paper to perform DOA estimation with co-prime arrays. In \cite{SuperRes} it was proved that the high frequency constant of a signal's spectrum can be perfectly recovered by sampling only the low end of its spectrum when the minimum distance among different spikes satisfies certain requirements. Robustness of this theory to noise is analyzed in \cite{SuperNoise}.  One merit of this theory is that it considers all the possible locations within the interested range, and thus does not suffer from model mismatch. Here we extend the mathematical theory of super resolution to DOA estimation with co-prime arrays under Gaussian noise. The noise structure resulting from the usage of co-prime arrays consists of a term with a known structure and another term consisting of quadratic combinations of Gaussian noise. Therefore, we modify the reconstruction method to fit this particular noise structure and show the robustness of our approach by analyzing the noise statistics. We also demonstrate theoretically that with $2M+N$ sensors in co-prime arrays, one can detect up to $\frac{MN}{2}$ sources.  Previous research \cite{corrGS} on identifiability using co-prime arrays was based on the idea of mutual coherence \cite{JAT06}. Although using mutual coherence can prove theoretically that by implementing co-prime arrays one can increase the number of sources being detected from $O(M+N)$ to $O(MN)$, this analysis based on coherence allows to go from $O(M+N)$ to $O(MN)$ only for very small values of the number of sources.

Source number detection is another main application of array processing. Various methods have been proposed over the years based on the eigenvalues of the signal space, such as the Akaike information criterion \cite{AIC}, the second order statistic of eigenvalues (SORTE) \cite{SORTE}, the predicted eigen-threshold approach \cite{ET}, and an eigenvector-based method that exploits the property of the variance of the rotational submatrix \cite{VTRS}. The authors of \cite{Han13} showed that among these methods, SORTE often leads to a better detection performance. We combine the SORTE method with spectrum reconstructed from DOA estimation to detect the number of sources. Through this source number detection, we referwhich reconstructed spikes are true detections and which are false alarms. 

The paper is organized as follows. In Section \ref{sec:DOA}, we introduce the DOA estimation model and explain how co-prime arrays can increase the degrees of freedom of the estimation system. In Section \ref{sec:SRT}, we extend super resolution theory to the application of co-prime arrays, and analyze the robustness of this extension by studying the statistics of the noise pattern in the model. We propose a numerical method to perform DOA estimation for co-prime arrays in Section \ref{sec:SDP}. We then extend this approach to detect the number of sources in Section \ref{sec:SND}. Section \ref{sec:NR} presents extensive numerical simulations to show the advantages of our methods in terms of estimation accuracy, degrees of freedom, and resolution ability. 

Throughout the paper, we use capital italic bold letters to represent matrices and operators, and lowercase italic bold letters to represent vectors. For a given matrix $\bs A$, $\bs A^*$ denotes the conjugate transpose matrix, $\bs A^{\mr T}$ denotes the transpose, and $\bs A^\mr{H}$ represents the conjugate matrix without transpose. We use $A_{mn}$ to denote the $(m,n)$th element of $\bs A$. We use $\otimes$ to denote the Kronecker product of two matrices. For a given operator $\bs F$, $\bs F^*$ denotes the conjugate operator of $\bs F$. Given vector $\bs x$, we use $\|\bs x\|_1$ and $\|\bs x\|_2$ to denote its $\ell_1$ and $\ell_2$ norms; $x_i$ and $x[i]$ are both used to represent the $i$th element of $\bs x$. Given a function $f$, $\|f\|_{L_1}, \|f\|_{L_2}, \|f\|_{L_\infty}$ are its $\ell_1, \ell_2, \ell_\infty$ norms. 

\section{Direction of Arrival Estimation and Co-Prime Arrays} \label{sec:DOA}
Consider a linear sensor with $L$ sensors which may be non-uniformly located. Assume that there are $K$ narrow band sources located at $\theta_1,\theta_2,\dots,\theta_K$ with signal powers $\sigma_1^2,\sigma_2^2,\dots,\sigma_K^2$. The steering vector for the $k$th source located at $\theta_k$ is $\bs a(\theta_k) \in \real^{L\times 1}$ with $l$-th element $e^{\bs j(2\pi/\lambda)d_l\sin(\theta_k)}$, in which $d_l$ is the location of the $l$th sensor and $\lambda$ is the wavelength. The data collected by all the sensors at time $t$ can be expressed as

\begin{equation}\label{eq:ULA}
\bs x(t)=\sum_{k=1}^K \bs a(\theta_k)s_k(t)+\bs \varepsilon(t)= \bs A \bs s(t)+\bs \varepsilon(t),
\end{equation}
for $t=1,\dots, T$, in which $\bs \varepsilon(t) \in \real^{L\times 1}$ is an i.i.d. white Gaussian noise $\mathcal{CN}(0,\sigma^2)$, $\bs A=[\bs a(\theta_1), \bs a(\theta_2)\dots, \bs a(\theta_K)] \in \real^{L\times K}$, and $\bs s(t)=[s_1(t),s_2(t),\dots,s_K(t)]^\mr T$ presents the source signal vector with $s_k(t)$ distributed as $\mathcal{CN}(0,\sigma_k^2)$. We assume that the sources are temporally uncorrelated. 

The correlation matrix among the $K$ sources can then be expressed as 
\begin{align}\label{eq:covar}
\bs R_{xx}=&E[\bs x(t)\bs x^*(t)] \notag\\
=&\bs A\bs R_{ss} \bs A^*+\sigma^2 \bs I\notag\\
=&\sum_{k=1}^K \sigma_k^2\bs a(\theta_k)\bs a^*(\theta_k)+\sigma^2\bs I,
\end{align}
in which $\bs R_{ss}$ is a $\real^{K\times K}$diagonal matrix with diagonal elements $\sigma_1^2,\sigma_2^2,\dots,\sigma_K^2$. After vectorizing the correlation matrix $\bs R_{xx}$, we have
 
\begin{equation}\label{eq:diff}
\bs z=\mr{vec}(\bs R_{\bs x\bs x})=\bs \Phi(\theta_1,\theta_2,\dots,\theta_K)\bs s+\sigma^2\bs 1_n,
\end{equation}
where 
$$\bs \Phi(\theta_1,\dots,\theta_K)=\bs A^*\odot \bs A=[\bs a(\theta_1)^\mr{H}\otimes\bs a(\theta_1),\dots, \bs a(\theta_K)^\mr{H}\otimes\bs a(\theta_K)].$$
The signal of interest becomes $\bs s=[\sigma_1^2,\sigma_2^2,\dots,\sigma_K^2]$, and $\bs 1_n=[\bs e_1^\mr T,\bs e_2^\mr T,\dots,\bs e_L^\mr T]^\mr T$, where $\bs e_i$ denotes a vector with all zero elements, except for the $i$th element, which equals to one. 

Comparing equations \eqref{eq:ULA} and \eqref{eq:diff}, we see that $\bs s$ behaves like a coherent source and $\sigma^2\bs 1_n$ becomes a deterministic noise term. The distinct rows in $\bs \Phi$ act as a larger virtual array with sensors located at $d_i-d_j$, with $1\leq i, j\leq L$. Traditional DOA estimation algorithms can be implemented to detect more sources when the structure of the sensor array is properly designed. Following this idea, nested arrays \cite{nested} and co-prime arrays \cite{coprime} were introduced, and then shown to improve the degrees of freedom from $O(N)$ to $O(N^2)$, and from $O(M+N)$ to $O(MN)$ respectively. In the following demonstration, we focus only on co-prime arrays; the results follow naturally for nested arrays.

\begin{figure}[h]
  \centering
  \includegraphics[width=0.45\textwidth]{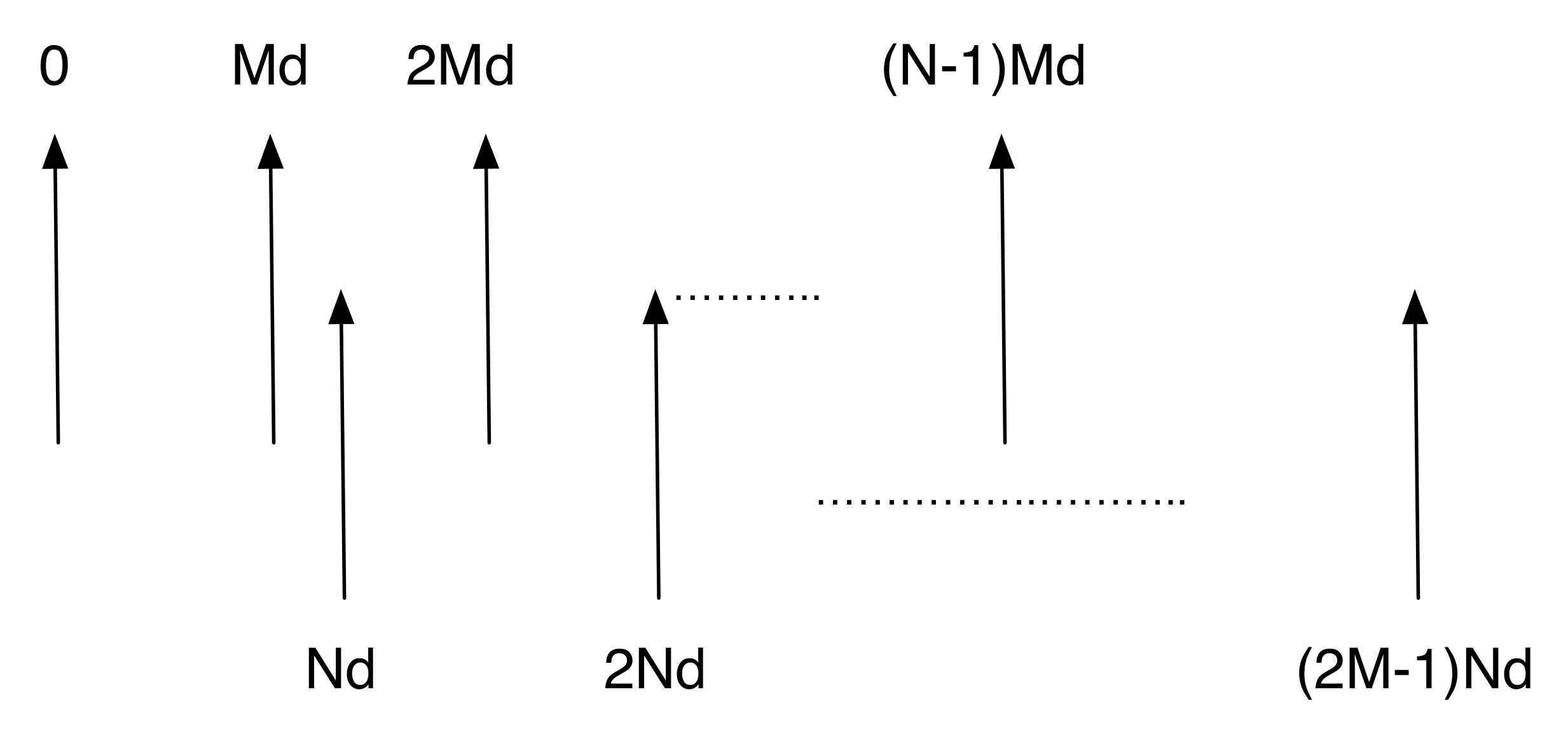}\\
  \caption{Geometry of Co-prime Arrays.}\label{fig:coprimemodel}
\end{figure}

Consider a co-prime array structure consist of two arrays with $N$ and $2M$ sensors respectively. The locations of the $N$ sensors are in the set $\{Mnd, 0\leq n \leq N-1\}$, and the locations of the $2M$ sensors are in the set $\{Nmd, 0\leq m\leq 2M-1\}$. Please note that first sensors of these two arrays are collocated. The geometry of these co-prime arrays is shown in Fig. \ref{fig:coprimemodel}. In this case the sensing matrix $\bs A \in \real^{(2M+N)\times K}$ has the same structure as that in \eqref{eq:ULA}. Indeed, the locations of the virtual sensors in \eqref{eq:diff} are given by the cross difference set $\{\pm(Mn-Nm)d, 0\leq n\leq N-1, 0\leq m \leq 2M-1\}$ and the two self difference sets. In order to implement spatial smoothing of MUSIC, we are interested in generating a consecutive range of virtual sensors. It was shown in \cite{comusic} that when $M$ and $N$ are coprime numbers, a consecutive range can be created from $-MNd$ to $MNd$, with $\{-MNd, -(MN-1)d,\dots,-2d,-d,d,2d,\dots,(MN-1)d,MNd\}$ taken from the cross difference set and $\{0d\}$ taken from any one of self difference sets.

By removing repeated rows of \eqref{eq:diff} and sorting the remaining rows from $-MNd$ to $MNd$, we have the linear model rearranged as
\begin{equation}\label{coULA}
\tilde{\bs z}=\tilde{\bs \Phi}\bs s+\sigma^2 \tilde {\bs w}.
\end{equation}
It is easy to verify that $\tilde {\bs w} \in \real^{(2MN+1)\times1}$ is a vector whose elements all equal to zero, except that the $(MN+1)$-th element equals to one. The matrix $\tilde{\bs \Phi} \in \real^{(2MN+1)\times K}$ is expressed as
$$\tilde{\bs \Phi}=\begin{bmatrix}
  e^{-\bs jMNd\frac{2\pi}{\lambda}\sin(\theta_1)} & \ldots & e^{-\bs jMNd\frac{2\pi}{\lambda}\sin(\theta_K)}\\
e^{-\bs j(MN-1)d\frac{2\pi}{\lambda}\sin(\theta_1)}   & \ldots & e^{-\bs j(MN-1)d\frac{2\pi}{\lambda}\sin(\theta_K)}\\
\vdots & \ddots & \vdots\\
 e^{\bs jMNd\frac{2\pi}{\lambda}\sin(\theta_1)} & \ldots & e^{\bs jMNd\frac{2\pi}{\lambda}\sin(\theta_K)}
\end{bmatrix},
$$
 which is the steering matrix for a uniform linear array (ULA) with $2MN+1$ sensors. Therefore, \eqref{coULA} can be regarded as a ULA detecting a coherent source $\bs s$ with deterministic noise term $\sigma^2 \tilde {\bs w}$. By applying MUSIC with spatial smoothing, the authors in \cite{comusic} showed that $O(MN)$ sources can be detected. 

\section{Direction of Arrival Estimation with Super Resolution Theory}\label{sec:SRT}
In this section we first assume that the signal model \eqref{eq:diff} is accurate, which means that the number of samples $T$ is infinity, and also that the noise power $\sigma^2$ is known a priori. The super resolution theory developed in \cite{SuperRes} can be implemented for co-prime arrays to demonstrate that we can detect up to $O(MN)$ sources as long as the distance between any two sources is on the order of $\frac{1}{MN}$. First we briefly introduce super resolution theory and extend the idea to the research of co-prime arrays. We then consider the case which the number of time samples $T$ is limited and demonstrate the robustness of super resolution via statistical analysis of the noise structure. 
 
\subsection{The Mathematical Theory of Super Resolution}
Super resolution seeks to recover high frequency details from the measurement of low frequency components. Mathematically, given a continuous signal $s(\tau)$ with $ \tau\in [0,1]$, the Fourier series coefficients are recorded as 
$$r(n)=\int_0^1 e^{-\bs j2\pi n\tau}s(d\tau), n=-f_c,-f_c+1,\dots,f_c.$$
Using the operator $\bs F$ to denote the low frequency measuring operator, we can write $\bs r=\bs F\bs s$, in which $\bs r=[r(-f_c),\dots,r(f_c)]^\mr T$ and $s=s(\tau), 0\leq \tau\leq 1$.

Suppose that the signal $s(\tau)$ is sparse, i.e., $s(\tau)$ is a weighted sum of several spikes:
\begin{equation}\label{spikes}
s(\tau)=\sum_{k=1}^K s_k\delta_{\tau_k},
\end{equation}
in which $s_k$ can be complex valued and $\tau_k \in [0,1]$ for all $k$. Then
\begin{equation}\label{supermodel}
r(n)=\sum_{k=1}^K s_ke^{-\mr j2\pi n\tau_k}, n=-f_c,-f_c+1,\dots,f_c. 
\end{equation}
Total variation minimization is introduced to encourage the sparsity in continuous signals $s(\tau)$, just as $\ell_1$ norm minimization produces sparse signals in the discrete space. Total variation for the complex measure $s$ is defined as
$$\|s\|_\mr {TV}=\sup\sum_{j=1}^\infty |s(B_j)|,$$
the supremum being taken over all partitions of the set $[0,1]$ into countable collections of disjoint measurable sets $B_j$. When $s$ takes the form in \eqref{spikes}, $\|s\|_\mr {TV}=\sum_{k=1}^K s_k$, which is the $\ell_1$ norm in the discrete case. The following convex optimization formula was proposed in \cite{SuperRes} to solve the super resolution problem:
\begin{equation}\label{opt:super}
\min_{\tilde{s}} \|\tilde{s}\|_\mr {TV} \quad \mr{s.t.} \quad \bs F \tilde{s}=\bs r .
\end{equation}
When the distance between any two $\tau_i$ and $\tau_j$ is larger than $2/f_c$, then the original sparse signal $s$ is the unique solution to the above convex optimization \cite{SuperRes}. The continuous optimization \eqref{opt:super} is solved via the following semidefinite programming \cite{SuperRes}:
\begin{align}
\max_{\bs u, \bs Q} \quad& \mr{Re} [\bs u^* \bs r]\notag \\
\mr{s.t.} \quad&
\begin{bmatrix}
  \bs Q & \bs u\\
 \bs u^*& 1
\end{bmatrix}\succeq 0,\\
& \sum_{i=1}^{2MN+1-j}\bs Q_{i,i+j}=\left\{
  \begin{array}{l l}
    1 & \quad j=0,\\
    0 & \quad j=1,2,\dots,2MN,\\
  \end{array} \right. \notag 
\end{align}
where $\bs Q \in \mathbb C^{(2MN+1)\times (2MN+1)}$ is an Hermitian matrix.

\subsection{DOA estimation with Super Resolution}
DOA estimation with co-prime arrays can be related to \eqref{supermodel} by a straightforward change of variables. Letting $\tau_k=\frac{d}{\lambda}(1-\sin(\theta_k))$ for all $k$, the linear model of \eqref{coULA} can be transformed into
\begin{align}\label{supercoprime}
r(n)=&e^{-\bs j2\pi n\frac{d}{\lambda}}(\tilde{z}_n-\sigma^2 w_n)=e^{-\bs j2\pi n\frac{d}{\lambda}}\sum_{k=1}^K s_ke^{\bs j 2\pi n \frac{d}{\lambda}\sin(\theta_k)} \notag
\\=& \sum_{k=1}^K s_ke^{-\mr j2\pi n \tau_k}, 
\end{align}
where $n=-MN,-MN+1,\dots,MN-1,MN. $ We use $\mathcal{T}=\{\tau_k, 1\leq k \leq K\}$ to denote the support set.

A theorem about the resolution and degrees of freedom for co-prime arrays can be directly derived using Theorem 1.2 in \cite{SuperRes}. Before introducing the theorem, we first define the minimum distance between any two sources as
$$\Delta (\bs \theta)= \min_{\theta_i,\theta_j,\theta_i \neq \theta_j} |\sin(\theta_i)-\sin(\theta_j)|.$$

\begin{theorem}
Consider a co-prime array consisting of two linear arrays with $N$ and $2M$ sensors respectively. The distances between two consecutive sensors are $Md$ for the first array and $Nd$ for the second array, where $M$ and $N$ are co-prime numbers, and $d\leq \frac{\lambda}{2}$. Suppose we have $K$ sources located at $\theta_1,\dots,\theta_K$. If the minimum distance follows the constraint that 
$$\Delta(\bs \theta) \geq \frac{2\lambda}{MNd},$$
then by solving the convex optimization \eqref{opt:super} with the signal model \eqref{supercoprime}, one can recover the locations $\theta_k$ for $k=1,\dots, K$ exactly. The maximum number of sources that can be detected is given by
$$K_\mr {max}=\frac{MNd}{\lambda}.$$
\end{theorem}

\noindent \textbf{Remark:}

\noindent With a traditional uniform linear array using $2M+N$ sensors, super resolution theory can detect up to $\frac{(2M+N)d}{2\lambda}$ sources when $\Delta(\bs \theta) \geq \frac{4\lambda}{(2M+N)d}$. With the utilization of co-prime arrays, the same number of sensors can detect $O(MN)$ sources as indicated by traditional MUSIC theory \cite{comusic}. As we will show in the numerical examples, implementing super resolution framework provides with a larger degrees of freedom and a finer resolution ability than those of MUSIC, since the spatial smoothing in the MUSIC reduces the obtained virtual array aperture.
 
\subsection{Noisy Model for Super Resolution}
In a realistic senario, the covariance matrix $\bs R_{xx}$ in \eqref{eq:covar} cannot be obtained exactly except unless the number of samples $T$ goes to infinity. Normally the covariance matrix is approximated by the following equation:
$$\hat{\bs R}_{xx}=\frac{1}{T}\sum_{t=1}^T \bs x(t)\bs x^*(t).$$
Subtracting the noise covariance matrix from both sides, we obtain
\begin{equation}\label{covarNoise}
\hat {\bs R}_{xx}-\sigma^2 \bs I=\bs A \bs R_{ss} \bs A^*+\bs E.
\end{equation}
Here $\bs R_{ss}$ is a diagonal matrix with $k$-th diagonal element 
$$\hat \sigma_k^2=\frac{1}{T}\sum_{t=1}^T s_k(t)s^*_k(t).$$
The $(m,n)$-th element in $\bs E$ is given as
\begin{align}\label{eq:Emn}
E_{mn}&=\frac{1}{T}\sum_{t=1}^T \sum_{i,j=1, i\neq j }^KA_{mi}A_{nj}^* s_i(t)s^*_j(t)\notag \\
&+\frac{1}{T}\sum_{t=1}^T\sum_{i=1}^K A_{mi} s_i(t) \varepsilon^*_n(t)+\frac{1}{T}\sum_{t=1}^T\sum_{i=1}^K\varepsilon_m(t)s_i^*(t) A_{ni}^* \notag\\
&+\frac{1}{T}\sum_{t=1}^T\varepsilon_m(t)\varepsilon^*_n(t)-\sigma^2 \bs I_{mn}, \quad 1\leq m, n \leq L.
\end{align}
Similar to the operation in \eqref{eq:diff},  vectorizing \eqref{covarNoise} leads to,
\begin{equation}\label{eq:noisevec}
\bs z=\mr{vec}(\bs \hat{R}_{\bs x\bs x})=\bs \Phi(\theta_1,\theta_2,\dots,\theta_K)\bs s+\sigma^2\bs 1_n+\bs e,
\end{equation}
where $\bs e$ is gained from vectorizing $\bs E$. For co-prime arrays, by removing repeated rows in \eqref{eq:noisevec}, and sorting them as consecutive lags from $-MNd$ to $MNd$, we get
\begin{equation}
\tilde{\bs z}=\tilde{\bs \Phi}\bs s+\sigma^2 \tilde {\bs w}+\tilde {\bs e}.
\end{equation} 
Please note that only one element from $\tilde{\bs e}$ corresponds to the diagonal element from $\bs E$. Here $\bs s=[\hat \sigma_1^2,\dots,\hat \sigma_K^2]^\mr T$. By applying the transformation technique in \eqref{supercoprime}, we have 
\begin{equation} \label{eq:noisesupermodel}
\bs r=\bs F \bs s+\bs e,
\end{equation}
where $e(n)=\tilde{e}(n) e^{-\bs j2\pi n\frac{d}{\lambda}}$. Thus we can formulate the following super resolution optimization problem, which considers the noise, as
\begin{equation}\label{opt:supernoise}
\min_{\bs s} \|\bs s\|_\mr {TV} \quad \mr{s.t.} \quad \|\bs F \bs s-\bs r\|_2 \leq \epsilon .
\end{equation}
The optimization can be solved via a semidefinite programming \cite{SuperNoise}:
\begin{align}
\max_{\bs u, \bs Q} \quad& \mr{Re} [\bs u^* \bs r]-\epsilon\|\bs u\|_2 \notag \\
\mr{s.t.} \quad&
\begin{bmatrix}
  \bs Q & \bs u\\
 \bs u^*& 1
\end{bmatrix}\succeq 0,\\
& \sum_{i=1}^{2MN+1-j}\bs Q_{i,i+j}=\left\{
  \begin{array}{l l}
    1 & \quad j=0,\\
    0 & \quad j=1,2,\dots,2MN.\\
  \end{array} \right. \notag 
\end{align}
Here $\bs Q \in \mathbb C^{(2MN+1)\times (2MN+1)}$ is an Hermitian matrix.

To derive the statistical behavior of each element in $\bs E$ we rely on two lemmas about the  concentration behavior of complex Gaussian random variables. Their proofs are based on the results from \cite{HBN10} and given in the Appendix. 
\begin{lemma}\label{lemma:xy}
Let $x(t)$ and $y(t), t=1,\dots,T$ be sequences of i.i.d., circularly-symmetric complex normal distributions with zero mean and variances equal to $\sigma_x^2$ and $\sigma_y^2$ respectively. That is $x(t) \sim \mathcal{CN}(0,\sigma_x^2)$ and $y(t) \sim \mathcal{CN}(0,\sigma_y^2)$. Then
$$\mr{Pr}\left (\left |\sum_{t=1}^T x(t)y^*(t)\right | \geq \epsilon \right ) \leq 8\exp \left (-\frac{\epsilon^2}{16\sigma_x\sigma_y(T\sigma_x\sigma_y+\frac{\epsilon}{4})}\right ).$$
\end{lemma}

\begin{lemma}\label{lemma:xx}
Let $x(t), t=1,\dots, T$ be a sequence of i.i.d., circularly-symmetric complex normal distribution with zero mean and variance equal to $\sigma_x^2$, i.e., $x(t) \sim \mathcal{CN}(0,\sigma_x^2)$. When $0\leq \epsilon \leq 4\sigma_x^2 T$, we obtain
\begin{align*}
\mr{Pr}\left ( \left |\sum_{t=1}^T x(t)x^*(t)-T\sigma_x^2\right | \geq \epsilon \right ) \leq 4\exp \left (-\frac{\epsilon^2}{16T\sigma_x^4}\right ).
\end{align*}
\end{lemma}

With these two concentration lemmas, the probability of $|E_{mn}|$ being larger than a constant can be upper bounded. For simplicity of analysis, in the rest of this paper we assume $\varepsilon\sim \mathcal{CN}(0, \sigma^2)$ and $s_i(t)\sim\mathcal{CN}(0, \sigma_s^2)$.

\begin{lemma}\label{lemma:Emn}
Let $E_{mn}$ be given in \eqref{eq:Emn}. Then for $m \neq n$ we have
\begin{align*}
\mr{Pr}(|E_{mn}|\geq \epsilon) \leq& 8\exp(-C_1(\epsilon)T)+16\exp(-C_2(\epsilon)T)\\
+&8\exp(-C_3(\epsilon)T).
\end{align*}
When $m=n$, we obtain
\begin{align*}
\mr{Pr}(|E_{mn}|\geq \epsilon) \leq& 8\exp(-C_1(\epsilon)T)+16\exp(-C_2(\epsilon)T)\notag \\
+&4\exp(-C_4(\epsilon)T),
\end{align*}
when $0 \leq \epsilon \leq 16 \sigma^2$. Here $C_1(\epsilon), C_2(\epsilon), C_3(\epsilon)$ and $C_4(\epsilon)$ are increasing functions of $\epsilon$.
\end{lemma}
\noindent \textbf{Proof:} We use $T_1, T_2$, and $T_3$ to denote the first three terms in \eqref{eq:Emn}. The last two terms are denoted by $T_4$. First we have
\begin{align*}
\mr{Pr}(|E_{mn}|\leq \epsilon) &\geq \mr{Pr}(\cap_{i=1}^4 |T_i| \leq \frac{\epsilon}{4})=1-\mr{Pr}(\cup_{i=1}^4 |T_i|\geq \frac{\epsilon}{4})\notag \\
& \geq 1-\sum_{i=1}^4 \mr{Pr}( |T_i|\geq \frac{\epsilon}{4}),
\end{align*} 
which leads to the inequality
\begin{equation}\label{ineq:Emn}
\mr{Pr}(|E_{mn}|\geq \epsilon) \leq \sum_{i=1}^4 \mr{Pr}( |T_i|\geq \frac{\epsilon}{4}).
\end{equation}
We also have
\begin{align} 
|T_1|&=\frac{1}{T}\sum_{t=1}^T \sum_{i,j=1, i\neq j }^KA_{mi}A_{nj}^* s_i(t)s^*_j(t)\notag \\
&\leq \frac{1}{T}\sum_{i,j=1,i\neq j}^K|A_{mi}A_{ni}^*|\left |\sum_{t=1}^Ts_i(t)s^*_j(t)\right |\notag\\
&\leq \frac{1}{T}\sum_{i,j=1,i\neq j}^K\left |\sum_{t=1}^Ts_i(t)s^*_j(t)\right |.
\end{align}
The last inequality follows from the fact that $|A_{mn}|\leq 1$ for all $m,n$. Thus
$$\mr{Pr}(|T_1|\geq \frac{\epsilon}{4}) \leq \mr{Pr}\left (\sum_{i,j=1,i\neq j}^K\left|\sum_{t=1}^Ts_i(t)s^*_j(t)\right|\geq \frac{\epsilon T}{4}\right ).$$
Then it is straightforward to find that 
$$\mr{Pr}(|T_1|\geq \frac{\epsilon}{4}) \leq \mr{Pr}\left (\left|\sum_{t=1}^Ts_{i_0}(t)s^*_{j_0}(t)\right|\geq \frac{\epsilon T}{4K(K-1)}\right ),$$
for some $i_0, j_0$ with $i_0 \neq j_0$. Using Lemma \ref{lemma:xy}
\begin{align}\label{eq:T1}
\mr{Pr}(|T_1|\geq \frac{\epsilon}{4}) \leq 8\exp(-C_1(\epsilon) T),
\end{align}
with $C_1(\epsilon)=\frac{\epsilon^2}{16\sigma_s^2K(K-1)(16\sigma_s^2K(K-1)+\epsilon)}.$ 

For the second term $T_2$, we have
\begin{align}
|T_2|&= \frac{1}{T}\sum_{t=1}^T\sum_{i=1}^K A_{mi} s_i(t) \varepsilon^*_n(t)\notag \\
&\leq \frac{1}{T}\sum_{i=1}^K|A_{mi}|\left|\sum_{t=1}^Ts_i(t)\varepsilon_n(t)^*\right | \leq \frac{1}{T}\sum_{i=1}^K\left |\sum_{t=1}^Ts_i(t)\varepsilon_n(t)^*\right |.
\end{align}
Following similar arguments as for $T_1$, we obtain that  
$$\mr{Pr}(|T_2|\geq \frac{\epsilon}{4}) \leq \mr{Pr}\left (\left|\sum_{t=1}^Ts_{i_0}(t)\varepsilon_n(t)^*\right|\geq \frac{\epsilon T}{4K}\right ).$$
Applying Lemma \ref{lemma:xy}, we have
\begin{equation}\label{eq:T2}
\mr{Pr}(|T_2|\geq \frac{\epsilon}{4}) \leq 8\exp(-C_2(\epsilon) T),
\end{equation}
with $C_2(\epsilon)=\frac{\epsilon^2}{16\sigma_s\sigma K(16\sigma_s\sigma K+\epsilon)}.$

For the third term, we have the same results as the second one, given as 
\begin{equation}\label{eq:T3}
\mr{Pr}(|T_3|\geq \frac{\epsilon}{4}) \leq 8\exp(-C_2(\epsilon) T).
\end{equation}
When $m \neq n$, the last term $T_4=\frac{1}{T}\sum_{t=1}^T\varepsilon_m(t)\varepsilon^*_n(t)$, and by Lemma \ref{lemma:xy}, 
\begin{equation}\label{eq:T41}
\mr{Pr}(|T_4|\geq \frac{\epsilon}{4}) \leq 8\exp(-C_3(\epsilon) T),
\end{equation}
with $C_3(\epsilon)=\frac{\epsilon^2}{16\sigma^2(16\sigma^2+\epsilon)}$. 
When $m=n$, the last term is given as $T_4=\frac{1}{T}\sum_{t=1}^T\varepsilon_m(t)\varepsilon^*_m(t)-\sigma^2$, thus the probability is bounded as 
\begin{equation}\label{eq:T42}
\mr{Pr}(|T_4|\geq \frac{\epsilon}{4}) \leq 4\exp(-C_4(\epsilon) T),
\end{equation}
where $C_4(\epsilon)=\frac{\epsilon^2}{256\sigma_\varepsilon^2}$ and $\epsilon \leq 16\sigma^2$. Applying the results from \eqref{eq:T1}, \eqref{eq:T2}, \eqref{eq:T3}, \eqref{eq:T41} and \eqref{eq:T42} to inequality \eqref{ineq:Emn}, we proves the remains. $\square$

In order to analyze the robustness of super resolution, a high resolution kernel is introduced in \cite{SuperNoise}  referred to as the Fej\'{e}r kernel. In our case it has a cut-off frequency $f_{h}>MN$ as is given by
\begin{align}\label{eq:Kh}
K_{h}(t)=&\frac{1}{f_h}\sum_{k=-f_h}^{f_h}(f_h+1-|k|)e^{\bs j2\pi kt}\notag \\
=&\frac{1}{f_h+1}\left (\frac{\sin(\pi(f_h+1)t)}{\sin({\pi t})}\right ).
\end{align}

Using the high resolution kernel $K_h(t)$ introduced in \eqref{eq:Kh}, we can show that by solving the convex optimization problem in \eqref{opt:supernoise} the high resolution details  of the original signal $s(\tau)$ can be recovered with high probability, even though the sample number $T$ is finite for co-prime arrays.
\begin{theorem}
Consider a co-prime array consisting of two linear arrays with $N$ and $2M$ sensors respectively. The distances between two consecutive sensors are $Md$ for the first array and $Nd$ for the second array, where $M$ and $N$ are co-prime numbers, and $d\leq \frac{\lambda}{2}$. Let $s(\tau)=\sum_{k=1}^K s_k\delta_{\tau_k}$. $T$ time sample points are collected for each receiver, by taking the transformation in \eqref{supercoprime} and solving the optimization \eqref{opt:supernoise} with $\bs s_\mr{opt}$ as the optimal function, we can show that 
$$\|K_{h}*(\bs s_\mr{opt}-\bs s)\|_{L_1} \leq C_0 \frac{f_h^2}{M^2N^2}\epsilon,$$
with probability at least $1-\alpha e^{-\gamma(\epsilon)T}$ when $\epsilon \leq 16\sqrt{2MN+1}\sigma^2$, where $\gamma(\epsilon)$ is a increasing function of $\epsilon$. Here $C_0$ is a positive constant number.
\end{theorem}

\noindent \textbf{Proof:} With the fact that $d\leq \frac{\lambda}{2}$, $\tau_k \in [0,1]$ for all $k$ after transformation \eqref{supercoprime}. It was shown in \cite{SuperNoise} that when the two conditions $\|\bs s_\mr{opt}\|_\mr{TV} \leq \|\bs s\|_\mr {TV}$ and $\|\bs F^*\bs F(\bs s_\mr{opt}-\bs s)\|_{L_1}\leq 2\epsilon$ hold, it suffices to obtain 
$$\|K_{h}*(\bs s_\mr{opt}-\bs s)\|_{L_1} \leq C_0 \frac{f_h^2}{M^2N^2}\epsilon.$$ 

In order to satisfy these conditions, the statistical behavior of $\bs e$ in \eqref{eq:noisesupermodel} is analyzed first. Using a similar argument to \eqref{ineq:Emn}, we have 
\begin{align}
\mr {Pr}(\|\bs e\|_2 \geq \epsilon) \leq& \sum_{n=-MN}^{MN} \mr{Pr}(|e(n)| \geq \frac{\epsilon}{\sqrt{2MN+1}}) \notag \\
= & \sum_{n=-MN}^{MN} \mr{Pr}(|\tilde{e}(n)| \geq \frac{\epsilon}{\sqrt{2MN+1}}).
\end{align}
The inequality follows from the fact that $|e(n)|=|\tilde{e}(n)|.$ $2MN$ elements of $\tilde{\bs e}$ are taken from $E_{mn}$ when $m \neq n$, and one element of $\tilde{\bs e}$ is taken from $E_{mn}$ when $m=n$. Therefore, by applying the results from Lemma \ref{lemma:Emn}, we can show that $\|\bs F\bs s-\bs r\|_2=\|\bs e\|_2 \leq \epsilon$ with a probability of at least $1-\alpha e^{-\gamma(\epsilon)T},$ and $\gamma(\epsilon)$ is a increasing function of $\epsilon$. The lemma requires that $\epsilon \leq 16\sqrt{2MN+1}\sigma^2.$

The first condition holds due to the optimization problem in \eqref{opt:supernoise}, and $\bs s$ is feasible with high probability. Furthermore,
\begin{align*}
\|\bs F^*\bs F(\bs s_\mr{opt}-\bs s)\|_{L_{1}} \leq & \|\bs F^*\bs F(\bs s_\mr{opt}-\bs s)\|_{L_{2}}=\|\bs F(\bs s_\mr{opt}-\bs s)\|_{L_{2}}\\
\leq & \|\bs F\bs s_\mr{opt}-\bs r\|_{L_{2}}+\|\bs F\bs s-\bs r\|_{L_{2}} \leq 2\epsilon.
\end{align*}  
The first inequality follows from the Cauchy-Schwarz inequality. Therefore the proof is complete. $\square$

\noindent \tb{Remark:}\\
$K_h$ defined in \eqref{eq:Kh} is a low pass filter with cut-off frequency $f_h>MN$. By convolving it with the reconstructed error $\bs s_\mr{opt}-\bs s$ we get the reconstruction error details up to the frequency $f_h$. By solving optimization \eqref{opt:supernoise}, using noisy measurement one can reconstruct the high frequency details of $\bs s$ with high probability. This probability goes to one exponentially as the number of samples $T$ goes to $\infty$. 

\section{DOA estimation via Semidefinite programming and Root Finding}\label{sec:SDP}
We now derive an optimization framework to reconstruct $\bs s$ for co-prime arrays. For DOA estimation the noise power $\sigma^2$ is normally unknown. Therefore, the optimization must be modified to include this effect. A more realistic optimization is reformulated as 
\begin{equation}\label{opt:prime}
\min_{\bs s, \sigma^2 \geq 0} \|\bs s\|_\mr {TV} \quad \mr{s.t.} \quad \|\bs F \bs s-\bs r-\sigma^2 \bs w\|_2 \leq \epsilon,
\end{equation}
in which $w_n=\tilde{w}_n e^{-\bs j2\pi n\frac{d}{\lambda}}$. The dual problem takes the form
\begin{align}\label{opt:dualorg}
&\max_{\bs u} \quad \mr{Re} [\bs u^* \bs r]-\epsilon\|\bs u\|_2 \notag \\
&\mr{s.t.}\quad \|\bs F^*\bs u\|_{L_{\infty}}\leq 1, \mr{Re} [\bs u^*\bs w]\leq 0. 
\end{align}
The derivation of the dual problem is given in the Appendix. Since $\bs u=\bs 0$ is a feasible solution, strong duality holds according to the general Slater's condition \cite{CVXopt}.

Due to the first constraint in \eqref{opt:dualorg}, the problem itself is still an infinite dimensional optimization. It was shown in \cite{SuperRes} that the first constraint can be recast as a semidefinite matrix constraint. Thus the infinite dimensional dual problem is equivalent to the following semidefinite programming (SDP):
\begin{align}\label{opt:dual}
\max_{\bs u, \bs Q} \quad& \mr{Re} [\bs u^* \bs r]-\epsilon\|\bs u\|_2 \notag \\
\mr{s.t.} \quad&
\begin{bmatrix}
  \bs Q & \bs u\\
 \bs u^*& 1
\end{bmatrix}\succeq 0, \quad \mr{Re}[\bs u^*\bs w] \leq 0,\\
& \sum_{i=1}^{2MN+1-j}\bs Q_{i,i+j}=\left\{
  \begin{array}{l l}
    1 & \quad j=0,\\
    0 & \quad j=1,2,\dots,2MN.\\
  \end{array} \right. \notag 
\end{align}
Here $\bs Q \in \mathbb C^{(2MN+1)\times (2MN+1)}$ is an Hermitian matrix. The optimization problem can be easily solved by using the CVX package \cite{CVXopt}. 

Solving \eqref{opt:dual} yields the optimal solution only for the dual problem. The following lemma is introduced to link the solutions of the primal and dual problems.

\begin{lemma}
Let $\bs s_\mr{opt}$ and $\bs u_\mr{opt}$ be the optimal solutions of the primal problem \eqref{opt:prime} and dual problem \eqref{opt:dual} respectively. Then 
$$\bs F^*\bs u_\mr {opt}(\tau)=\mr{sgn}(\bs s_\mr{opt}(\tau))$$
for all $\tau$ such that $\bs s_\mr{opt}(\tau) \neq 0.$
\end{lemma}

\noindent \textbf{Proof:} Let $\sigma_\mr{opt}^2$ be the noise power estimated in the primal problem. Since strong duality holds, we have
\begin{align*}
\|\bs s_\mr {opt}\|_\mr{TV}&=\mr{Re} \langle \bs r,\bs u_\mr{opt}\rangle-\epsilon \|\bs u_\mr {opt}\|_2\notag\\
&=\mr{Re}\langle \bs r-\bs F\bs s_\mr{opt}-\sigma_\mr{opt}^2 \bs w,\bs u_\mr{opt}\rangle-\epsilon \|\bs u_\mr{opt}\|_2\notag \\
&+\mr{Re}\langle \bs F\bs s_\mr{opt}+\sigma_\mr{opt}^2 \bs w,\bs u_\mr{opt}\rangle \notag \\
&\leq \mr{Re}\langle \bs F\bs s_\mr{opt}+\sigma_\mr{opt}^2 \bs w,\bs u_\mr{opt}\rangle \leq \mr{Re}\langle \bs F\bs s_\mr{opt},\bs u_\mr{opt}\rangle. 
\end{align*}
The first inequality follows from the Cauchy-Schwarz inequality and the fact that $\| \bs r-\bs F\bs s_\mr{opt}-\sigma_\mr{opt}^2 \bs w\|_2\leq \epsilon$. The second inequality results from $\mr {Re}[\bs u_\mr{opt}^*\bs w]\leq 0$. Because $\|\bs F^*\bs u_\mr {opt}\|_{L_\infty}\leq 1$, we have  $\|\bs s_\mr {opt}\|_\mr{TV} \geq \mr{Re}\langle \bs s_\mr{opt},\bs F^*\bs u_\mr{opt}\rangle$. Therefore $\|\bs s_\mr {opt}\|_\mr{TV} = \mr{Re}\langle \bs s_\mr{opt},\bs F^*\bs u_\mr{opt}\rangle$ holds and we have the desired result needed to satisfy this equality. $\square$

The support set $\mathcal{T}$ can be estimated by root-finding based on the trigonometric polynomial $1-|\bs F^*\bs u(\tau)|^2=0$. Let $\mathcal{T}_\mr{est}$ denote the estimation of the support sets, and use $\tau_\mr{est}[i]$ to denote elements in $\mathcal{T}_\mr{est}$ with $1\leq i\leq K_\mr{est}.$ A matrix $\bs F_\mr {est} \in \mathbb{C}^{(2MN+1)\times K_\mr{set}}$ can be formulated, with measurement $\bs r$ expressed as
\begin{equation}\label{linear_dis}
\bs r=\bs F_\mr {est} \bs s_0+\sigma^2 \bs w+\bs e,
\end{equation}
in which $\bs s_0 \in \real^{K_\mr {est}}$ and 
$$\bs F_\mr{est}=\begin{bmatrix}
  e^{-\bs jMNd 2\pi \tau_\mr{est}[1]} & \ldots & e^{-\bs jMNd 2\pi \tau_\mr{est}[K_\mr{est}]}\\
  e^{-\bs j(MN-1)d 2\pi \tau_\mr{est}[1]} & \ldots & e^{-\bs j(MN-1)d 2\pi \tau_\mr{est}[K_\mr{est}]}\\
\vdots & \ddots & \vdots\\
   e^{\bs jMNd 2\pi \tau_\mr{est}[1]} & \ldots & e^{\bs jMNd 2\pi \tau_\mr{est}[K_\mr{est}]}
\end{bmatrix}.
$$

Due to the the numerical issue in the root finding process, the cardinality of $\mathcal{T}_\mr{set}$ is normally larger than the cardinality of  $\mathcal{T}$, i.e., $K_\mr{est}\geq K$. It is possible in some cases that $K_\mr{set} \geq 2MN+1$, which leads to an ill conditional linear system \eqref{linear_dis}. Sparsity can then be exploited on this signal $\bs s_0$. A convex optimization in the discrete domain can be formulated as 
\begin{equation}\label{opt:disprime}
\min_{\bs s_0, \sigma^2 \geq 0} \|\bs s_0\|_1 \quad \mr{s.t.} \quad \|\bs F_\mr{est} \bs s_0-\bs r-\sigma^2 \bs w\|_2 \leq \epsilon_d.
\end{equation}
The $\epsilon_d$ in \eqref{opt:disprime} is normally chosen to be larger than $\epsilon$ in \eqref{opt:prime} since the noise level is expected to be higher in \eqref{linear_dis} due to inevitable error introduced in the root finding process. Assuming that the optimization solution of \eqref{opt:disprime} is $\bs s^\mr{est} \in \real^{K_\mr{set}}$, the estimation of $\bs s$ in the continuous domain can be represented as 
$$\bs s_\mr{opt}=\sum_{i=1}^{K_\mr{est}}s_\mr{est}[i] \delta_{\tau_\mr {est}[i]}.$$

\section{Extension: Source Number Detection}\label{sec:SND}
Traditional source number detection for array processing is typically performed by exploiting eigenvalues from the sample covariance matrix. For coprime arrays, this covariance matrix can be obtained by performing spatial smoothing on $\tilde{\bs z}$. The same idea can also be implemented on the sparse signal recovered from the previous section.  Ideally, after sorting its elements in a descending order, the signal $\bs s_\mr{est}$ reconstructed from \eqref{opt:disprime} should follow
\begin{align*}
s_\mr{est}[1]^2 & \geq s_\mr{est}[2]^2 \geq \dots s_\mr{est}[K]^2 \\
&\geq s_\mr{est}[K+1]^2= \dots = s_\mr{est}[K_\mr{est}]^2=0. 
\end{align*}

The SORTE algorithm can be applied to this series. The difference of the elements from $\bs s_\mr{set}$ is 
\begin{equation*}
\gr s_\mr{est}[i]=s_\mr{est}[i]^2-s_\mr{est}[i+1]^2, \text{for } i=1,\dots,K_\mr{est}-1.
\end{equation*}
The gap measure in SORTE is given as 
\begin{equation}\label{eq:sorte}
\mr {SORTE}(i)=\left\{
  \begin{array}{l l}
    \frac{\mr{var}[i+1]}{\mr{var}[i]} & \mr{var}[i] \neq 0,\\
    +\infty & \mr{var}[i]=0,\\
  \end{array} \right. i=1,\dots,K_\mr{est}-2, 
\end{equation}
where 
\begin{equation}
\mr{var}[i]=\frac{1}{K_\mr{est}-i} \sum_{m=i}^{K_\mr{est}-1}\left(\gr s_\mr{est}[m]-\frac{1}{K_\mr{est}-i}  \sum_{n=i}^{K_\mr{est}-1} \gr s_\mr{est}[n]\right)^2.
\end{equation}

The number of the sources can be determined by following the criteron
$$\hat{K}=\mr {argmin}_i \quad \mr{SORTE}(i).$$
It only works when $K_\mr{set} > 2$ due to the definition of $\mr{SORTE}(i)$ in \eqref{eq:sorte}. When $K_\mr{est} \leq 2$, since $\mathcal{T}_\mr{est}$ is obtained from the rooting finding process based on the continuous sparse recovery, we simply let $\hat K =K_\mr{est}$. We will refer to this continuous sparse recovery based SORTE as CSORTE.

\section{Numerical Results} \label{sec:NR}
In this section, we present several numerical examples to show the merits of implementing super resolution techniques on co-prime arrays. We consider a co-prime array with 11 sensors. One set of sensors is located at positions $[0,3,6,9,12]d$, and the second set of sensors is located at positions $[0,5,10,15,20,25]d$, where $d$ is taken as half of the wavelength. The first sensors from both sets are collocated. It is easy to show that the correlation matrix generates a virtual array with lags from $-17d$ to $17d$. We compare the continuous sparse recovery (CSR) techniques with MUISC and also with  the discrete sparse recovery method (DSR) considering grid mismatches \cite{ZT13_SPL}. In \cite{ZT13_SPL}, a LASSO formulation was used to perform the DOA estimation. Here we implement an equivalent form of LASSO, i.e., Basis pursuit, to perform the comparison. The MUSIC method in this simulation follows the spatial smoothing technique in \cite{comusic}. For the discrete sparse recovery method, we take the grid from $-1$ to $1$, with step size $0.005$ for $\sin(\theta)$. The noise levels $\epsilon$ in the optimization formulas are chosen by cross validation. We consider $15$ narrow band signals located at $\sin(\bs \theta)=[-0.8876, -0.7624, -0.6326, -0.5096, -0.3818, -0.2552,\\ -0.1324, -0.0046, 0.1206, 0.2414, 0.3692, 0.4972, 0.6208,\\ 0.7454, 0.8704]$. We show that the continuous sparse recovery method yields better results in terms of detection ability, resolution, and estimation accuracy.
 
\subsection{Degrees of Freedom}
\begin{figure}[h]
  \centering
  \includegraphics[width=0.5\textwidth]{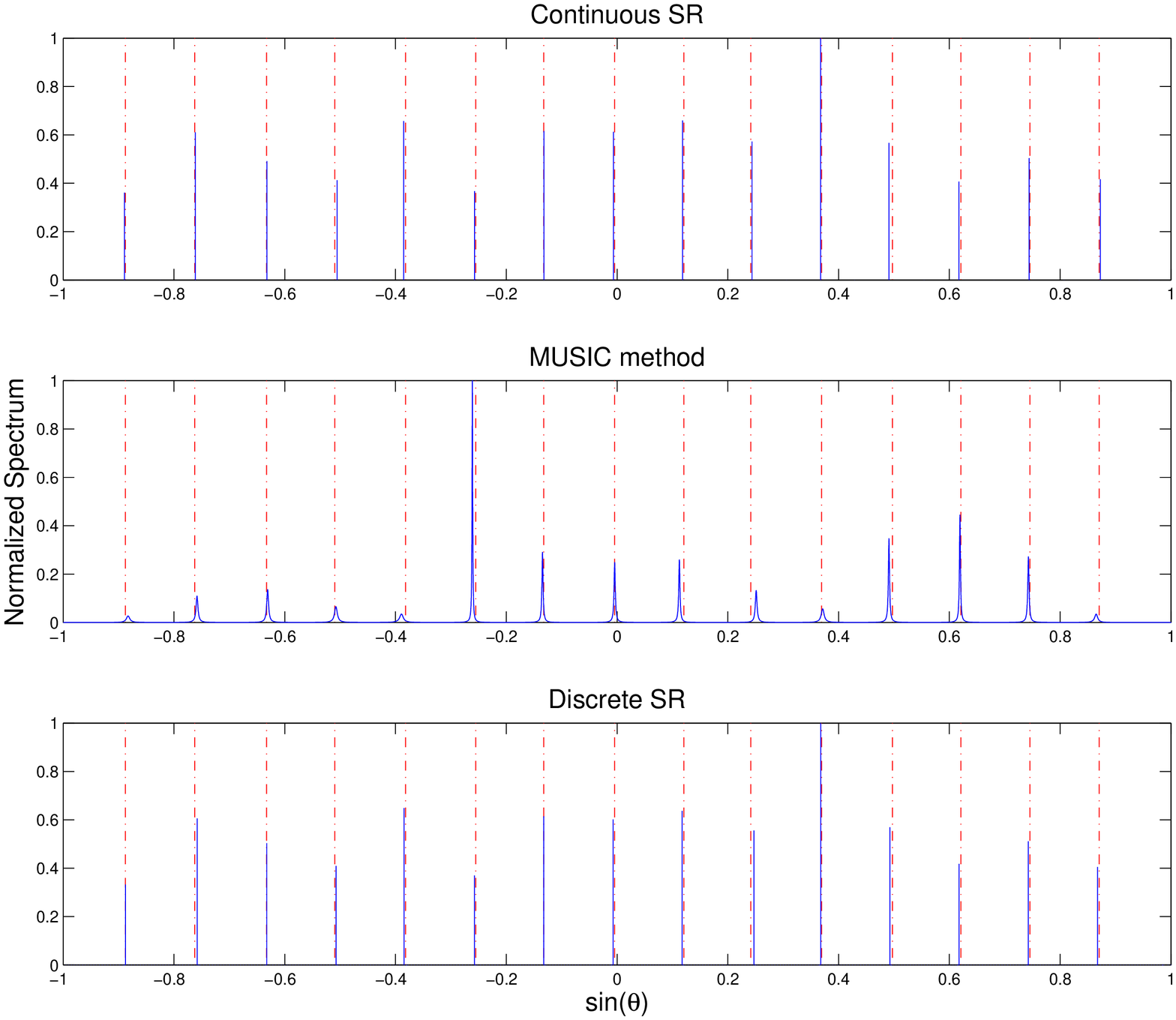}\\
  \caption{Normalized spectra for CSR, MUSIC, and DSR, with $T=500$ and SNR=$-10$dB.}\label{fig:example}
\end{figure}

In this first numerical example, we verify that continuous sparse recovery increases the degrees of freedom to $O(MN)$ by implementing the coprime arrays' structure. The $\epsilon$ for CSR is taken as $5$, and $\epsilon_d$ is taken as $10$ while DSR uses $\epsilon=10$. The number of time samples is $500$ and the SNR is chosen to be $-10$dB. In Fig. \ref{fig:example}, we use a dashed line to represent the true directions of arrival. The CPU time for running CSR was $7.30$ seconds. DSR took $7.82$ seconds, while MUSIC algorithm only used $0.81$ seconds. For MUSIC we implement a root MUSIC algorithm to estimate the location of each source, and the number of sources is assumed to be given. The average estimation errors for CSR, DSR, and root MUSIC are $0.23\%, 0.26\%$, and $0.42\%$ respectively. We can see that all the three methods achieve $O(MN)$. In the following subsection, we test the estimation accuracy of these three methods via Monte Carlo simulations.

\subsection{Estimation Accuracy}
In this section, we compare the continuous sparse recovery method with the MUSIC algorithm and also the discrete sparse recovery method via Monte Carlo simulations. Since traditional MUSIC does not yield the DOA of each source directly, we consider the Root MUSIC algorithm instead. For simplicity, we will still refer it as MUSIC in this section. The number of sources is assumed to be known for the MUSIC algorithm in this simulation, while sparse methods do not assume this a priori. $\epsilon$ and $\epsilon_d$ are chosen to be $5$ and $10$ in this simulation, while discrete SR uses $\epsilon=10$.

\begin{figure}[h]
  \centering
  \includegraphics[width=0.5\textwidth]{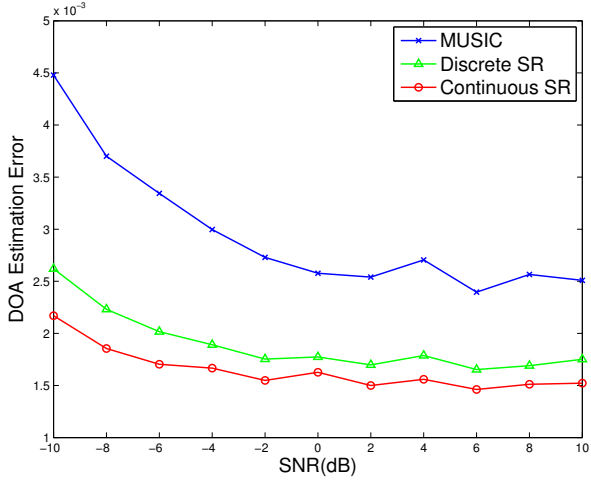}\\
  \caption{DOA estimation errors for CSR, MUSIC, and DSR, with $T=500$.}\label{fig:AccSNR}
\end{figure}

Figure \ref{fig:AccSNR} shows the DOA estimation error with respect to changing SNR after 50 Monte Carlo simulations. The estimation error is calculated based on the sine function of the DOAs. The average CPU times for running CSR, DSR and MUSIC are $6.93$s, $9.30$s, and $1.46$s respectively. We can see that CSR performs better than DSR uniformly with less computing time. Both sparse recovery methods achieve better DOA estimation accuracy than MUSIC. The accuracy of DSR can be further improved by taking finer grid with a smaller stepsize. However, this will slow down the computing of DSR further.

\begin{figure}[h]
  \centering
  \includegraphics[width=0.5\textwidth]{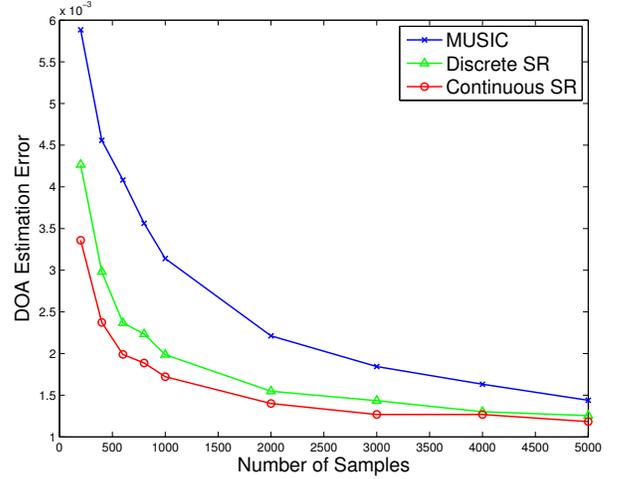}\\
  \caption{DOA estimation error for CSR, MUSIC, and DSR, with SNR=$-10$ dB.}\label{fig:AccT}
\end{figure}

In Fig. \ref{fig:AccT} we show that with a changing number of snapshots the proposed CSR also exhibits better estimation accuracy than either DSR or MUSIC. The average CPU times for running CSR, DSR and MUSIC are $6.50$s, $7.91$s, and $1.43$s respectively. The performance of MUSIC and DSR approach the performance of CSR when the number of snapshots approaches $5000$. We can see that implementing CSR can save sampling time by taking a small number of snapshots to achieve the same estimation accuracy as the MUSIC algorithm. The parameters $\epsilon$ and $\epsilon_d$ are equal to $5$ and $10$ in this simulation.

\subsection{Source Number Detection Performance Comparsion}
In this section, we compare the source number detection performance of the proposed CSORTE with that of traditional SORTE applied to the covariance matrix. The SNR is set to $0$dB while the number of snapshots is $3000$. We range the number of sources from $11$ to $17$. Since this co-prime array structure yields consecutive lags from $-17d$ to $17d$, $17$ is the maximum number of sources that can be detected theoretically via techniques based on the covariance matrix.

\begin{figure}[h]
  \centering
  \includegraphics[width=0.5\textwidth]{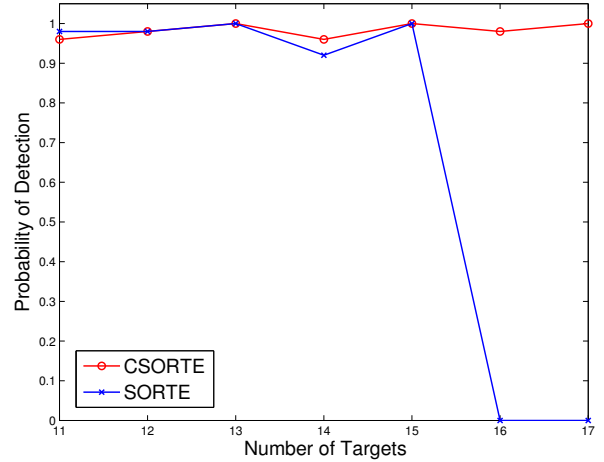}\\
  \caption{Source number detection using CSORTE and SORTE, with SNR=$0$ dB, $T=3000$.}\label{fig:SN}
\end{figure}

Figure \ref{fig:SN} shows the probability of detection with respect to the number of sources after $50$ Monte Carlo simulations. In the CSR, $\epsilon$ is chosen to be $5\sigma$, and $\epsilon_d$ is set to be $2 \epsilon$. When the number of sources is less than $15$, CSORTE and SORTE  yield comparable result. However, SORTE fails after the number of sources is larger than $15$, while CSORTE gives stable performance and also exhibits perfect detection even when the number of sources reaches the theoretical limit of $17$. Discrete sparse recovery can also be combined with SORTE to perform source number detection. However, the detection accuracy is jeopardized by the spurious signal from the reconstructed signals using DSR. Therefore SORTE based on DSR is not included here. This simulation shows that the sparsity based method offers more degrees of freedom than the subspace based method.

\subsection{Resolution Ability}

Finally we compare the resolution abilities of the MUSIC algorithm and the proposed continuous sparse recovery method. We show that  CSR is capable of resolving very closely located signals. In the first simulation, two sources are closely located at $-32^\circ$ and $-30^\circ$.The value of $\epsilon$ is chosen to be $0.7 \sigma$ and $\epsilon_d$ is set to be $2 \epsilon$ in the CSR, where $\sigma$ is the noise power. 

\begin{figure}[h]
  \centering
  \includegraphics[width=0.5\textwidth]{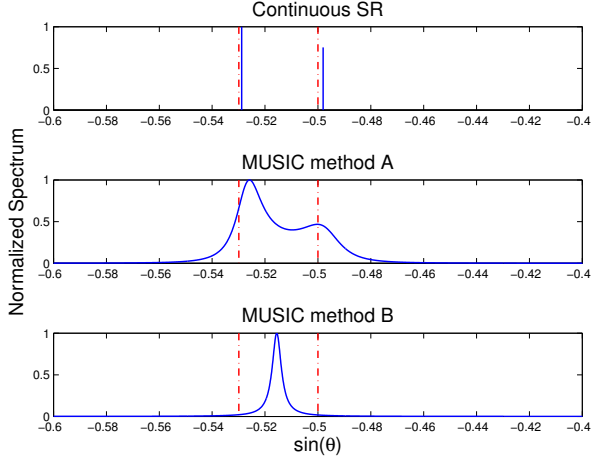}\\
  \caption{Source number detection using CSR and the MUSIC algorithm, with SNR=$0$ dB, $T=500$.}\label{fig:SN0db}
\end{figure}

\begin{figure}[h]
  \centering
  \includegraphics[width=0.5\textwidth]{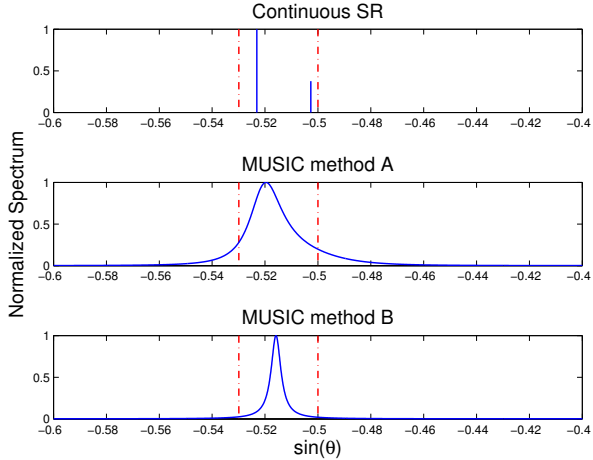}\\
  \caption{Source number detection using CSR and MUSIC algorithm, with SNR=$-5$ dB, $T=500$.}\label{fig:SN}
\end{figure}

Figure \ref{fig:SN0db} shows a numerical example when the SNR is $0$ dB and the number of snapshots is $500$. Normalized spectra are plotted for three methods. MUSIC method A is the MUSIC algorithm with the assumption that the number of sources is known while MUSIC method B is the MUSIC method relying on traditional SORTE to provide the estimated number of sources. We can see that MUSIC method B fails to resolve these two targets because the traditional SORTE fails to estimate the number of sources correctly. CSR resolves the two sources successfully even though a priori information about the number of sources is not assumed to be given. In Fig. \ref{fig:SN}, we lower the SNR to $-5$ dB, and we notice that even given the number of sources, the MUSIC algorithm fails to resolve the two closely located sources while CSR resolves successfully. Next we conduct a simulation based on Monte Carlo runs to compare the resolution ability of the CSORTE and the traditional SORTE algorithm. 

\begin{figure}[h]
  \centering
  \includegraphics[width=0.5\textwidth]{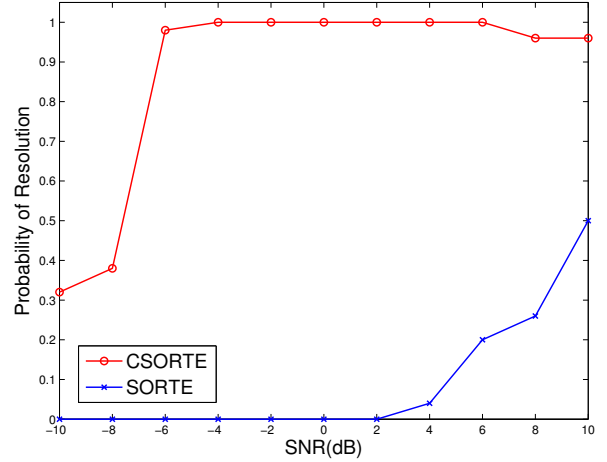}\\
  \caption{Comparison of resolution performance of  CSORTE and SORTE, with $T=2000$.}\label{fig:resSN}
\end{figure}

Figure \ref{fig:resSN} shows the resolution performance in detecting two sources located at  $-32^\circ$ and $-30^\circ$, using CSORTE and SORTE methods after $50$ Monte Carlo runs. The parameter $\epsilon$ is chosen to be $0.7 \sigma$, and $\epsilon_d$ is set to be $2 \epsilon$ in the CSR. We can see that CSORTE outperforms the traditional SORTE when detecting the two closely located sources.

\section{Conclusions and Future Work}
In this work, we extended the mathematical theory of super resolution to the topic of DOA estimation using co-prime arrays. A primal-dual approach was utilized to transform the original infinite dimensional optimization to a solvable semidefinite program. After estimating the candidate support sets by solving the semidefinite program, a small scale sparse recovery problem can be solved efficiently. The robustness of the proposed super resolution approach was verified by performing statistical analysis of the noise inherit to co-prime arrays processing. A source number detection algorithm was then proposed by combining the existing SORTE algorithm with the reconstructed spectrum from convex optimization. Via numerical examples, we showed that the proposed method achieves a more accurate DOA estimation while providing more degrees of freedom, and also exhibits a more powerful resolution ability than the traditional MUSIC algorithm with spatial smoothing.

Although implementing the continuous sparse recovery method saves sampling time in obtaining a certain estimation accuracy compared with MUSIC, one shortcoming of the proposed sparse method is that solving the semidefinite program is more time consuming than the MUSIC algorithm. Fast algorithm development could be an interesting topic for future work. It is also of interest to develop a systematic way to choose $\epsilon$ and $\epsilon_d$ in the optimization formulas. One major assumption made by current co-prime arrays research is that sources are uncorrelated. Incoporating correlations among sources is also an important topic for future work.

\appendix

\noindent \tb{Proof of Lemma \ref{lemma:xy}:}

First we have
\begin{align*}
\sum_{t=1}^T x(t)y^*(t)&=\sum_{t=1}^T \mr {Re}[x(t)]\mr {Re}[y(t)]+ \sum_{t=1}^T \mr {Im}[x(t)]\mr {Im}[y(t)] \\
&-\bs j\sum_{t=1}^T \mr {Re}[x(t)]\mr {Im}[y(t)]+ \bs j\sum_{t=1}^T \mr {Im}[x(t)]\mr {Re}[y(t)].
\end{align*}
According to the same procedure used in the proof of lemma \ref{lemma:Emn}, we have
\begin{align*}
&\mr{Pr}\left (\left |\sum_{t=1}^T x(t)y^*(t)\right |\geq \epsilon \right )\\
\leq &\mr{Pr}\left (\left |\sum_{t=1}^T \mr {Re}[x(t)]\mr {Re}[y(t)]\right |\geq \frac{\epsilon}{4}\right)\\
+&\mr{Pr} \left(\left|\sum_{t=1}^T \mr {Im}[x(t)]\mr {Im}[y(t)]\right|\geq \frac{\epsilon}{4} \right) \\
+&\mr{Pr}\left(\left|\sum_{t=1}^T \mr {Re}[x(t)]\mr {Im}[y(t)]\right|\geq \frac{\epsilon}{4}\right)\\
+& \mr{Pr}\left(\left|\sum_{t=1}^T \mr {Im}[x(t)]\mr {Re}[y(t)]\right|\geq \frac{\epsilon}{4}\right).
\end{align*}  
Using the lemma 6 from \cite{HBN10}, we finish the proof of lemma \ref{lemma:xy} $\square$.

Before the next proof, we need to show that the square sum of i.i.d Gaussian random variables concentrate around the sum of the variance. It utilizes the result in lemma 7 from \cite{HBN10}.
\begin{lemma}\label{lemma:myxx}
Let $x(t), t=1,\dots,T$ be a sequence of i.i.d. normal distributions with zero mean and variance equal to $\sigma^2$, i.e., $x(t) \sim \mathcal{N}(0,\sigma^2)$. Then 
$$\mr{Pr}\left (|\sum_{t=1}^T x(t)^2-T\sigma^2|\geq \epsilon \right) \leq 2\exp(-\frac{\epsilon^2}{16\sigma^4T})$$
when $0\leq \epsilon \leq 4\sigma^2 T$.
\end{lemma} 
\noindent \tb{Proof:} From the results in \cite{HBN10}, we know that for any positive $c$, we have two asymmetric bounds as 
$$\mr{Pr}\left(\sum_{t=1}^Tx(t)^2-T\sigma^2 \geq 2\sigma^2\sqrt{Tc}+2\sigma^2c\right) \leq \exp(-c),$$
$$\mr{Pr}\left(\sum_{t=1}^Tx(t)^2-T\sigma^2 \leq -2\sigma^2\sqrt{Tc}\right) \leq \exp(-c).$$
When $0\leq c\leq T$, we obtain
$$\mr{Pr}\left(\sum_{t=1}^Tx(t)^2-T\sigma^2 \geq 4\sigma^2\sqrt{Tc}\right) \leq \exp(-c),$$
$$\mr{Pr}\left(\sum_{t=1}^Tx(t)^2-T\sigma^2 \leq -4\sigma^2\sqrt{Tc}\right) \leq \exp(-c).$$
Combing the above two inequalities, we get
$$\mr{Pr}\left(|\sum_{t=1}^Tx(t)^2-T\sigma^2| \geq 4\sigma^2\sqrt{Tc}\right) \leq 2\exp(-c),$$
which yields the result by replacing $4\sigma^2\sqrt{Tc}$ with $\epsilon$ while maintaining $0\leq c\leq T$.

\noindent \tb{Proof of Lemma \ref{lemma:xx}:}

First we have the equation
\begin{align*}
\sum_{t=1}^T x(t)x^*(t)-T\sigma_x^2&=\sum_{t=1}^T \mr {Re}[x(t)]^2+ \sum_{t=1}^T \mr {Im}[x(t)]^2-T\sigma_x^2.
\end{align*}
Likewise, we obtain
\begin{align*}
&\mr{Pr}(|\sum_{t=1}^T x(t)x(t)^*-T\sigma_x^2|\geq \epsilon)\\
=&\mr{Pr}(|\sum_{t=1}^T \mr {Re}[x(t)]^2-\frac{T\sigma_x^2}{2}|\geq \frac{\epsilon}{2})\\
+& \mr{Pr}(|\sum_{t=1}^T \mr {Im}[x(t)]^2-\frac{T\sigma_x^2}{2}|\geq \frac{\epsilon}{2}).
\end{align*}
With lemma \ref{lemma:myxx}, we have the desired result for lemma \ref{lemma:xx}.

\noindent \tb{Derivation of the dual problem in Section \ref{sec:SDP}}

By introducing by variable $\bs z\in \mathbb{C}^{2MN+1}$, the original primal problem is equivalent to the following optimization:
\begin{align*}
&\min_{\bs s, \sigma^2 \geq 0, \bs z} \|\bs s\|_\mr {TV} \\
 \mr{s.t.} \quad \|\bs z \|_2\leq &\epsilon,\quad \bs z=\bs F \bs s-\bs r-\sigma^2 \bs w.
\end{align*}
With the Lagrangian multiplier $v \in \real$ and $\bs u \in \mathbb{C}^{2MN+1}$, the Lagrangian function is given as
\begin{align*}
L(\bs s,\bs z,\sigma^2,\bs u,v)=&\|\bs s\|_\mr {TV}+v(\|\bs z\|_2-\epsilon)\\
+&\mr {Re}[\bs u^*(\bs r-\bs F\bs s-\sigma^2\bs w-\bs z)].
\end{align*}
The dual function is given as 
\begin{align*}
&g(\bs u,v)=\mr{Re}[\bs u^*\bs r]-v\epsilon\\
+&\inf_{\bs s,\bs z,\sigma^2 \geq 0}\{\|\bs s\|_\mr{TV}-\mr{Re}[\bs u^*\bs F\bs s]-\sigma^2\mr{Re}[\bs u^*\bs w]+v\|\bs z\|_2-\bs u^*\bs z\}.
\end{align*}
The Lagrangian multipliers $\bs u$ and $v$ in the domain of the dual function have to satisfy the following three constraints:
$$\|\bs F^*\bs u\|_{L_\infty}\leq 1, \mr{Re}[\bs u^*\bs w]\leq 0, v\frac{\bs z}{\|\bs z\|_2}=\bs u.$$
From the third constraint, we have $v=\|\bs u\|_2$. Therefore, we obtain the dual problem stated in \eqref{opt:dual}.
%------------------------------------------------------------------------
%   References
% ------------------------------------------------------------------------
\bibliographystyle{IEEEtran}
\bibliography{IEEEabrv,coprime}

\printnomenclature
\bibliographystyle{plain}

\end{document}